\documentclass[prb,twocolumn,floatfix,showpacs]{revtex4}
\usepackage{graphicx,color}
\usepackage{amsmath,amssymb,bm}

\newcommand{\mc}{\mathcal}
\newcommand{\la}{\Lambda}
\newcommand{\tn}{\textnormal}
\newcommand{\cprb}[3]{Phys.~Rev.~B {\bf #1}, #2 (#3)}
\newcommand{\cprl}[3]{Phys.~Rev.~Lett.~{\bf #1}, #2 (#3)}
\newcommand{\cnjp}[3]{New J.~Phys.~{\bf #1}, #2 (#3)}
\newcommand{\cjp}[3]{J.~Phys.: Condensed Matter {\bf #1}, #2 (#3)}
\newcommand{\cnature}[3]{Nature {\bf #1}, #2 (#3)}
\newcommand{\cbook}[2]{\textit{#1} (#2)}

\definecolor{darkred}{rgb}{0.90,0,0}
\definecolor{darkgreen}{rgb}{0,0.60,.2}
\definecolor{darkblue}{rgb}{0,0,1}
\definecolor{grey}{cmyk}{0,0,0,0.25}
\definecolor{orange}{cmyk}{0,0.6,0.8,0}

\begin{document}
\title{\boldmath Supercurrent and multiple singlet-doublet phase transitions\\ of a quantum dot Josephson junction inside an Aharonov-Bohm ring}
\author{C.\ Karrasch and V.\ Meden}
\address{Institut f\"ur Theoretische Physik A and JARA -- Fundamentals of Future Information Technology, RWTH Aachen University, 52056 Aachen, Germany}

\begin{abstract}
We study a quantum dot Josephson junction inside an Aharonov-Bohm environment. The geometry is modeled by an Anderson impurity coupled to two directly-linked BCS leads. We illustrate that the well-established picture of the low-energy physics being governed by an interplay of two distinct (singlet and doublet) phases is still valid for this interferometric setup. The phase boundary depends, however, non-monotonically on the coupling strength between the superconductors, causing the system to exhibit re-entrance behavior and multiple phase transitions. We compute the zero-temperature Josephson current and demonstrate that it can become negative in the singlet phase by virtue of the Coulomb interaction $U$. As a starting point, the limit of large superconducting energy gaps $\Delta=\infty$ is solved analytically. In order to tackle arbitrary $\Delta<\infty$ and $U>0$, we employ a truncated functional renormalization group scheme which was previously demonstrated to give quantitatively reliable results for the quantum dot Josephson problem.
\end{abstract}

\maketitle

\section{Introduction}
\label{sec:intro}

The low-energy behavior of quantum dot Josephson junctions is governed by an interplay of superconductivity and the Kondo effect. The physics emerging from the competition of these correlation phenomena was discussed decades ago in the context of magnetic impurities inside superconducting metals.\cite{oldwork1,oldwork2,oldwork3} If the Kondo energy scale $T_K$ is much larger than the superconducting gap $\Delta$, local magnetic moments are screened by virtue of the Kondo effect. This causes Cooper pairs to break, and the ground state of the system becomes a Kondo rather than a BCS singlet. In the opposite limit $T_K\ll\Delta$, Kondo screening is disturbed due to the superconducting gap at the Fermi energy, and the ground state describes free magnetic moments. At temperature $T=0$, a first order level-crossing quantum phase transition from a non-magnetic singlet to a degenerate (so-called magnetic) ground state is observed if $\Delta/T_K$ increases.

In recent years, the rise of nanotechnology allowed for sandwiching quantum dots between superconducting electrodes and for measuring the equilibrium Josephson current as a function of well-controlable microscopic parameters.\cite{exp1,exp2,exp3,exp4,exp5,exp6,exp7,exp8,expnovotny,expdaenen,helene} This was the motivation to theoretically re-investigate the interplay between superconductivity and Kondo physics in the context of (Anderson-like) models which feature all parameters (and not only $\Delta/T_K$) necessary to describe the experimental quantum dot Josephson junction.\cite{gm,ra,vecino,oguri1,choi,novotny,se,oguri2,bcspaper} Both the phase boundary between the (Kondo) singlet and the (magnetic) doublet phases and the supercurrent were calculated using reliable many-particle methods to tackle the vital Coulomb interaction $U$.\cite{oguri1,oguri2,bcspaper} In particular, is was demonstrated that the critical supercurrent as a function of the quantum dot energy $\epsilon$ can be obtained in good agreement with experimental data.\cite{helene}

If a quantum dot is placed in one arm of a non-superconducting closed Aharonov-Bohm geometry, signatures of the Fano effect can be experimentally observed in mesoscopic systems.\cite{expfano1,expfano2,expfano3} In addition, the interferometric setup allows for extracting physical properties which cannot be accessed by measurements on the isolated dot (such as the transmission phase).\cite{expphase1,expphase2,expphase3} Both situations were investigated theoretically using Anderson-like impurity models as well as appropriate many-body methods in order to obtain a physical understanding consistent with the observed data.\cite{fanokondo,phaselapses} In contrast, no interferometric experiments on quantum dots within a superconducting environment have been performed so far. However, in consideration of the rapid progress in nano science it is reasonable to assume that experimental (e.g., transport) data on such setups will become available fairly soon.

As mentioned above, both the singlet-doublet phase transition and the supercurrent of the quantum dot Josephson junction were extensively investigated theoretically, most times employing the single impurity Anderson model with BCS source and drain leads.\cite{gm,ra,vecino,oguri1,choi,novotny,se,oguri2,bcspaper} In contrast, the Aharonov-Bohm situation where both superconductors are in addition directly linked by a hopping matrix element $t_d$ has only been partly investigated. Zhang used a slave-boson mean-field approach to compute the supercurrent for the (singlet) situation where the Kondo temperature $T_K$ is larger than the energy gap $\Delta$.\cite{zhang} However, the author fails to obtain correct results in the analytically-solveable non-interacting case $U=0$, rendering his results questionable.\cite{zhangU0} The opposite (doublet) situation with $T_K<\Delta$ was studied by Osawa, Kurihara, and Yokoshi.\cite{oky} They employ, however, a Hartree-Fock framework which cannot account properly for Kondo correlations, the latter being a vital ingredient for the problem at hand. Most surprisingly, both works do not at all address the question whether the general picture of the existence of singlet and doublet low-energy states survives if the superconductors are connected directly, and, if so, how the ground state actually depends on the physical parameters (particularly $t_d$) of the system. It is the first aim of this paper to clarify this issue and to demonstrate that the $T=0$ `phase boundary' is affected non-monotonically by a finite coupling $t_d>0$, causing the system to exhibit re-entrance behavior and multiple singlet-doublet transitions. Secondly, we present reliable results for the zero-temperature Josephson current $J$ for arbitrary system parameters (not focussing on a specific regime of $\Delta/T_K$\cite{commentkondo}) and particularly illustrate that $J$ can become negative in the singlet phase.\cite{zhangJ} Our starting point is the so-called atomic limit $\Delta=\infty$ which can be treated analytically even in presence of finite Coulomb correlations. In order to address arbitrary $\Delta<\infty$ and $U>0$, we employ the functional renormalization group (FRG). By comparison with accurate data obtained from the numerical renormalization group framework, it was illustrated that this (after truncation) approximate method succeeds both qualitatively and quantitatively in producing the phase boundary and supercurrent for the simple quantum dot Josephson junction ($t_d=0$).\cite{bcspaper} In addition, we will demonstrate that for $t_d>0$ the FRG scheme benchmarks excellently against the analytic result at $\Delta=\infty$, thereby altogether providing a reliable tool to study the problem at hand.

This paper is organized as follows. In Sec.~\ref{sec:model}, we introduce the single impurity Anderson model with directly linked BCS superconducting leads and compute the associated non-interacting impurity Green function. The limit $\Delta=\infty$ is solved analytically in Sec.~\ref{sec:method.atom}, and Sec.~\ref{sec:method.frg} is devoted to a short introduction of the FRG framework. The phase boundary of the singlet-doublet level-crossing phase transition as well as the Josephson current are discussed in Secs.~\ref{sec:pb} and \ref{sec:j}, respectively. We conclude our paper with a short summary (Sec.~\ref{sec:outlook}).

\begin{figure}[t]
\includegraphics[width=0.7\linewidth,clip]{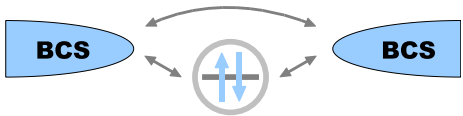}
\caption{(Color online) The interferometric quantum dot Josephson junction considered in this paper.}
\label{fig:geometry}
\end{figure}

\section{Model}
\label{sec:model}

In order to describe the geometry depicted in Fig.~\ref{fig:geometry}, we introduce the standard BCS and Anderson impurity Hamiltonian
\begin{alignat}{5}\label{eq:model.h1}
&H^\tn{dot} && =  \sum_\sigma \epsilon d^\dagger_\sigma d_\sigma + U \left(d^\dagger_\uparrow d_\uparrow-\frac{1}{2}\right)\left(d^\dagger_\downarrow d_\downarrow-\frac{1}{2}\right)~,\nonumber \\
&H^\tn{lead}_{s=L,R} && = \sum_{k\sigma}\epsilon_{sk}c^\dagger_{sk\sigma}c_{sk\sigma} 
- \Delta\sum_k\left(e^{i\phi_s}c^\dagger_{sk\uparrow}c^\dagger_{s-k\downarrow} + \tn{H.c.}\right)\tag{1a}\nonumber
\end{alignat}
as well as the coupling terms
\begin{alignat}{5}\label{eq:model.h2}
&H^\tn{coup}_{s=L,R} && = -\frac{t}{\sqrt{N}}\sum_{k\sigma}c_{sk\sigma}^\dagger d_\sigma + \tn{H.c.}~,\nonumber \\
&H^\tn{direct} &&= -\frac{t_d}{N}\sum_{k_1k_2\sigma}c^\dagger_{Lk_1\sigma}c_{Rk_2\sigma}+\tn{H.c.}~.\tag{1b}\nonumber
\end{alignat}
Both $c_{sk\sigma}$ and $d_\sigma$ denote usual fermionic annihilation operators. The quantum dot is characterized by a gate voltage $\epsilon$ and a local Coulomb repulsion $U$ between spin-up and spin-down electrons. For simplicity, we assume that the left ($s=L$) and right ($s=R$) BCS leads have equal superconducting energy gaps $\Delta$ while exhibiting a finite phase difference $\phi=2\phi_L=-2\phi_R$. Both leads are locally coupled to each other and to the quantum dot by (real) hopping amplitudes $t_d$ and $t$, respectively.\setcounter{equation}{1}

As a first step to solve the quantum many-particle problem associated with Eq.~(\ref{eq:model.h1},\ref{eq:model.h2}), it is useful to compute the non-interacting Green function of the quantum dot. This is achieved straight-forwardly using the equation-of-motion technique.\cite{oky,pinkbook} One obtains
\begin{equation}\label{eq:model.g0}\begin{split}
\mc G^0(z) & =
\begin{pmatrix}
\langle\langle d_\uparrow d_\uparrow^\dagger\rangle\rangle_z &
\langle\langle d_\uparrow d_\downarrow\rangle\rangle_z \\
\langle\langle d_\downarrow^\dagger d_\uparrow^\dagger\rangle\rangle_z &
\langle\langle d_\downarrow^\dagger d_\downarrow\rangle\rangle_z
\end{pmatrix}_{U=0} \\[1ex]
& =\frac{1}{z-\epsilon\tau_3+t\tau_3A_L(z)+t\tau_3A_R(z)}~.
\end{split}\end{equation}
Here, $\tau_i$ denote the Pauli matrices, and $A_s$ determines the lead `self-energy':
\begin{equation}
A_{s=L,R}(z)= t\frac{t_dg_s(z)\tau_3g_{\bar s}(z)\tau_3-g_s(z)\tau_3}{1-t_d^2g_s(z)\tau_3g_{\bar s}(z)\tau_3}~,
\end{equation}
with the definition $\bar L=R$, $\bar R=L$ as well as the implicit understanding (here and in the following) that the inverse is multiplied from the left. The local Green function $g_s(z)$ of the isolated BCS leads is given by
\begin{equation}
g_s(z) = -\frac{\pi\rho}{\sqrt{\Delta^2-z^2}}
\begin{pmatrix} z & -\Delta e^{i\phi_s} \\ -\Delta e^{-i\phi_s} & z \end{pmatrix}~.
\end{equation}
We have assumed the local density of states $\rho=\sum_k\delta(\epsilon-\epsilon_{sk})/N$ and thus the hybridization energy
\begin{equation}
\Gamma=2\pi\rho t^2
\end{equation}
to be constant, implementing the so-called wide-band limit.

Following the arguments presented in Ref.~\onlinecite{bcspaper}, one can show that the Josephson current $J_s=i\langle[H,N_s]\rangle$ (with $N_{s=L,R}$ being the particle number operator of the left and right lead, respectively) can be computed from the exact expression (taking $\hbar=1$ and the electron charge $e=1$ in the following)\nopagebreak
\begin{equation}\label{eq:model.j}
J_s = 2T\sum_{i\omega}\tn{Im}\,\tn{Tr}\,\left[t\mc G_{sd}(i\omega)-t_d\mc G_{s\bar s}(i\omega)\right]~,
\end{equation}
where $T$ is the temperature of the system, and $\mc G_{sd}(z)$ and $\mc G_{s\bar s}(z)$ denote the interacting dot-lead and lead-lead Green function, respectively. The first and second term of Eq.~(\ref{eq:model.j}) can naturally be regarded as the impurity and direct contribution to the supercurrent. Employing the equation-of-motion technique and generalizing the resulting relations by virtue of the Dyson equation for $U\neq0$, both $\mc G_{sd}(z)$ and $\mc G_{s\bar s}(z)$ can be expressed in terms of the interacting dot Green function $\mc G(z)$ as follows:
\begin{equation}\begin{split}
\mc G_{sd}(z) & = A_s(z)\mc G(z)~,\\
\mc G_{s\bar s}(z) & = A_s(z)\mc G(z) A_{\bar s}^\dagger(z^*)
-\frac{t_dg_s(z)\tau_3g_{\bar s}(z)}{1-t_d^2g_s(z)\tau_3g_{\bar s}(z)\tau_3}~.
\end{split}\end{equation}
At $\Gamma=0$, the right-hand side of Eq.~(\ref{eq:model.j}) can be evaluated analytically up to second order in $t_d$, leading to the sinusoidal law $J\sim\sin(\phi)$ which describes the ordinary Josephson junction.\cite{pinkbook}

Employing a gauge transformation, one can show that the supercurrent obtained from Eq.~(\ref{eq:model.j}) fulfills $J_L=-J_R=J$, provided that $\mc G(z)$ is given exactly.\cite{bcspaper} In the present paper, we focus exclusively on the situation of fully symmetric superconducting leads (featuring equal energy gaps $\Delta=\Delta_L=\Delta_R$ as well as equal hybridization strengths $\Gamma=2\Gamma_L=2\Gamma_R$) and can thus refrain (since $J_L=-J_R$ trivially holds) from addressing the issue of current conservation within the approximate functional RG approach introduced below.\cite{bcspaper}

\section{Solution strategies}
\label{sec:method}
In order to study the quantum many-particle problem implicated by the Hamiltonian of Eq.~(\ref{eq:model.h1},\ref{eq:model.h2}), we proceed as follows. First, we demonstrate that in the limit of large BCS gaps $\Delta=\infty$ one can derive an analytic expression for the phase boundary describing the singlet-doublet phase transition (Sec.~\ref{sec:method.atom}). In order to tackle the Coulomb correlation $U$ at arbitrary $\Delta<\infty$, we introduce a truncated functional renormalization group scheme (Sec.~\ref{sec:method.frg}). By comparison with numerical RG data, it was previously shown that the latter provides an accurate tool to calculate the phase boundary as well as the supercurrent for the (non-interferometric) quantum dot Josephson junction.\cite{bcspaper}

\subsection{\boldmath Analytic treatment of the limit $\Delta=\infty$}
\label{sec:method.atom}
For $t_d=0$, it was previously demonstrated (see Refs.~\onlinecite{oguri2,bcspaper}) that the limit of large superconducting gaps $\Delta=\infty$ allows for an analytic treatment even in presence of finite Coulomb correlations $U$. This is still possible for the Aharonov-Bohm situation. Namely, at $\Delta=\infty$ the non-interacting dot Green function [Eq.~(\ref{eq:model.g0})] becomes
\begin{equation}
\left[\mc G^0(z)\right]^{-1}=
\begin{pmatrix} z-\tilde\epsilon & \tilde\Delta \\ \tilde\Delta & z+\tilde\epsilon \end{pmatrix}~,
\end{equation}
where the effective parameters $\tilde\epsilon$ and $\tilde\Delta$ are given by
\begin{equation}\begin{split}
\tilde\epsilon & = \epsilon+\Gamma\frac{\tilde t_d\cos(\phi)+\tilde t_d^3}{1+2\tilde t_d^2\cos(\phi)+\tilde t_d^4}~, \\
\tilde\Delta & = \Gamma\cos(\phi/2)\frac{1+\tilde t_d^2}{1+2\tilde t_d^2\cos(\phi)+\tilde t_d^4}~,
\end{split}\end{equation}
and $\tilde t_d=\pi\rho t_d$. Including the interacting part, the problem is thus equivalent to diagonalizing the effective two-particle Hamiltonian
\begin{equation}\begin{split}
H_\tn{eff} = &~ \tilde\epsilon d^\dagger_\uparrow d_\uparrow+\tilde\epsilon d^\dagger_\downarrow d_\downarrow
-\tilde\Delta\left(d_\uparrow^\dagger d_\downarrow^\dagger+d_\downarrow d_\uparrow\right) \\
& +U \left(d^\dagger_\uparrow d_\uparrow-\frac{1}{2}\right)\left(d^\dagger_\downarrow d_\downarrow-\frac{1}{2}\right)~.
\end{split}\end{equation}
This can be achieved straight-forwardly by virtue of a Bogolyubov transformation.\cite{pinkbook} It turns out that the ground state of the system is either non-degenerate (a singlet which at sufficiently large $U$ can be thought of featuring Kondo screening and broken Cooper pairs) or doubly degenerate (a `magnetic' doublet generally associated with a free spin), illustrating that this well-known picture is still valid in presence of a finite coupling $t_d>0$. By comparison of the corresponding many-particle energies one can show that the level crossing and thus the zero-temperature `phase transition' is determined by the implicit equation
\begin{equation}\label{eq:method.atom.pb}
U^2 = 4\tilde\epsilon^2+4\tilde\Delta^2~.
\end{equation}
Since $\cos(\phi)$ can become negative for $0\leq\phi\leq\pi$, the right-hand side of Eq.~(\ref{eq:method.atom.pb}) is not necessarily a monotonic function of the bare parameters $\epsilon/\Gamma$ and $t_d/\Gamma$, immediately indicating re-entrance behavior and multiple singlet-doublet phase transitions. This will be discussed in detail in Sec.~\ref{sec:pb}.

\begin{figure*}[t]
\includegraphics[width=0.85\linewidth,clip]{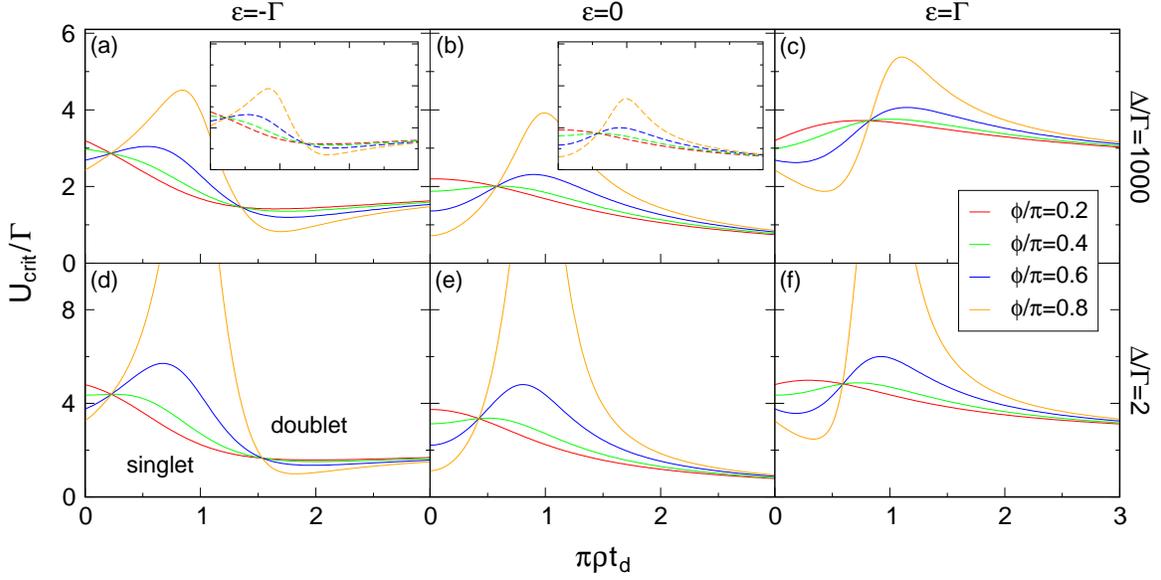}
\caption{(Color online) The critical interaction strength $U_\tn{crit}$ as a (non-monotonic) function of the direct coupling $t_d$ for different BCS gaps $\Delta$ and impurity energies $\epsilon$, altogether characterizing the singlet-doublet level-crossing phase transition of the Aharonov-Bohm quantum dot Josephson junction. Solid lines where obtained from the FRG approach introduced in Sec.~\ref{sec:method.frg}, dashed lines display the analytic result derived in the limit $\Delta=\infty$ [see Eq.~(\ref{eq:method.atom.pb})]. The phase difference between the left and right superconducting leads is given by $\phi=0.2\pi$, $0.4\pi$, $0.6\pi$, and $0.8\pi$ (from top to bottom at $t_d=0$). The axis of the insets are scaled the same as the axis of the corresponding main part.}
\label{fig:pb1}
\end{figure*}

\begin{figure}[b]
   \includegraphics[height=5.8cm,clip]{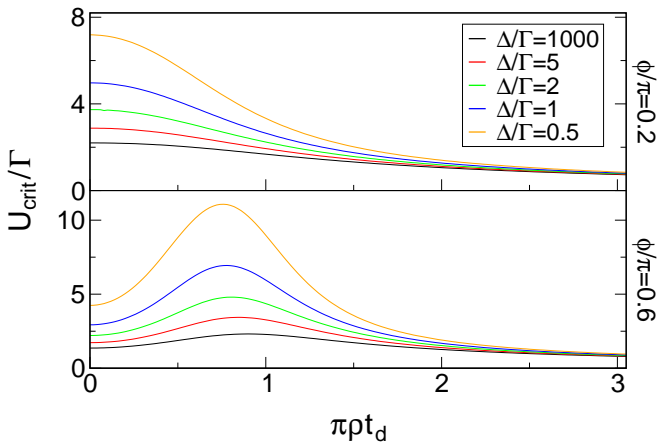}
   \caption{(Color online) The same as Fig.~\ref{fig:pb1}, but for fixed $\epsilon=0$ and different $\Delta/\Gamma=1000$, $5$, $2$, $1$, and $0.5$ (from bottom to top). The phase boundary always resembles the analytic form derived in the limit $\Delta=\infty$ [Eq.~(\ref{eq:method.atom.pb})], only the size of the doublet phase shrinks monotonously with the BCS energy gap.}
   \label{fig:pb2}
\end{figure}

\begin{figure*}[tb]
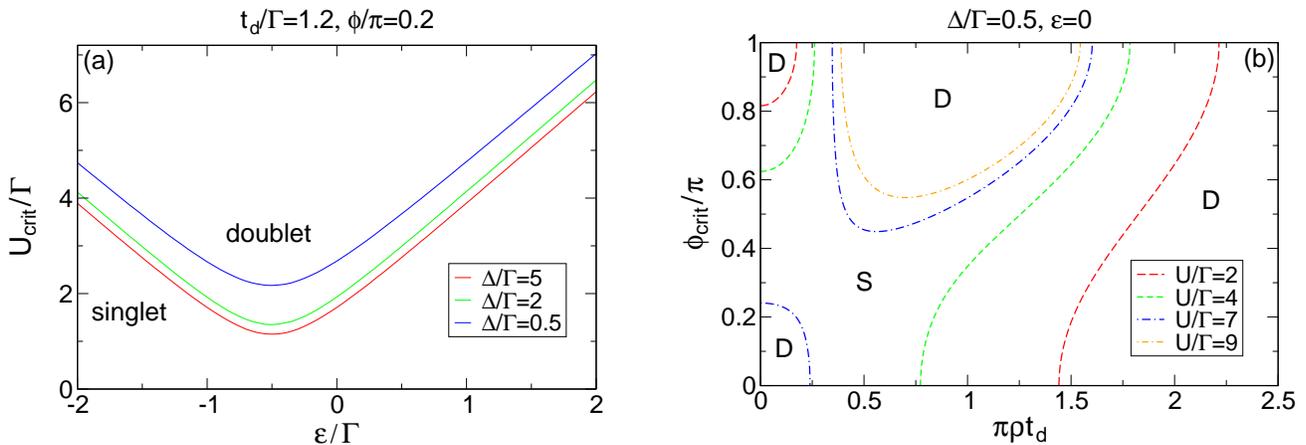

   \includegraphics[height=5.8cm,clip]{PB3.eps}\hspace*{0.05\linewidth}
   \includegraphics[height=5.8cm,clip]{PB4.eps}
   \caption{(Color online) (a) FRG results for the critical interaction $U_\tn{crit}$ as an (almost quadratic) function of the impurity energy $\epsilon$ for $\Delta/\Gamma=5$, $2$, and $0.5$ (from bottom to top). (b) Critical phase difference $\phi_\tn{crit}$ separating the singlet (S) from the doublet (D) phase as a function of the direct coupling $t_d$. The displayed behavior is similar to analytic result derived at $\Delta=\infty$.}
   \label{fig:pb3}
\end{figure*}

\subsection{Functional renormalization group approach}
\label{sec:method.frg}
The functional renormalization group is one implementation of Wilson's general RG idea for interacting many-particle systems.\cite{salmhofer} It starts with introducing an energy cutoff $\la$ into the non-interacting Green function of the system under consideration. Here, we choose a multiplicative infrared cutoff $\Theta(|\omega|-\la)$ in Matsubara frequency space. By taking the derivative of many-particle vertex functions (such as the self-energy) with respect to the cutoff parameter $\la$, one obtains an infinite hierarchy of flow equations, and subsequent integration from $\la=\infty$ down to the cutoff-free system $\la=0$ leads to an in principle exact solution of the many-particle problem. In practice, however, the infinite hierarchy needs to be truncated, rendering the FRG an approximate method. In this paper, we employ a truncation scheme that keeps the flow equations for the self-energy and the two-particle vertex evaluated at zero external frequencies. The resulting approximation to both quantities is frequency-independent, contains at least all terms up to order $U$, and can be computed numerically with minor effort. It was demonstrated in recent works that this truncated FRG scheme successfully describes correlation effects (e.g., aspects of Kondo physics) in quantum impurity systems.\cite{phaselapses,dotsystems} In particular, comparison with numerical RG reference data showed that both the singlet-doublet phase transition and the Josephson current of a (non-interferometric) quantum dot Josephson junction can be computed reliably using this framework.\cite{bcspaper}

The FRG flow equations for the diagonal and anomalous part of the self energy $\Sigma^\la$ and $\Sigma_\Delta^\la$ associated with the Hamiltonian of Eq.~(\ref{eq:model.h1},\ref{eq:model.h2}) can be obtained by a slight generalization of the derivation presented in Ref.~\onlinecite{bcspaper}. They are given by
\begin{equation}\label{eq:method.frg.flow1}\begin{split}
\dot\Sigma^\la = & \frac{U^\la}{\pi}\,\tn{Re}\left[\tilde{\mc G}^\la_{22}(i\la)\right]~, \\
\dot\Sigma_\Delta^\la = & -\frac{U^\la}{2\pi}\left[\tilde{\mc G}^\la_{12}(i\la)+\tilde{\mc G}^\la_{12}(-i\la)\right]~,
\end{split}\end{equation}
and the flow equation of the effective interaction $U^\la$ reads
\begin{equation}\label{eq:method.frg.flow2}\begin{split}
\dot U^\la  = & \frac{(U^\la)^2}{\pi}\,\tn{Re}\,
\Big[ \tilde{\mc G}^\la_{12}(i\la)\tilde{\mc G}^\la_{21}(i\la) + \tilde{\mc G}^\la_{12}(i\la)\tilde{\mc G}^\la_{21}(-i\la)\\
& -\tilde{\mc G}^\la_{11}(i\la)\tilde{\mc G}^\la_{22}(i\la) - \tilde{\mc G}^\la_{11}(i\la)\tilde{\mc G}^\la_{22}(-i\la) \Big]~.
\end{split}\end{equation}
We have defined the matrix $\tilde{\mc G}^\la(i\omega)$ via
\begin{equation}
\Big[\tilde{\mc G}^\la(i\omega)\Big]^{-1} = \Big[\mc G^0(i\omega)\Big]^{-1} -
\begin{pmatrix}
\Sigma^\la & \Sigma_\Delta^\la \\ (\Sigma_\Delta^\la)^* & -\Sigma^\la
\end{pmatrix}~.
\end{equation}
The initial conditions to the coupled differential equations (\ref{eq:method.frg.flow1},\ref{eq:method.frg.flow2}) read $\Sigma^{\la\to\infty}=0$, $\Sigma_\Delta^{\la\to\infty}=0$, and $U^{\la\to\infty}=U$, and one can carry out a numerical integration using standard Runge-Kutta routines in order to obtain the frequency-independent FRG approximation $\Sigma=\Sigma^{\la=0}$ and $\Sigma_\Delta=\Sigma_\Delta^{\la=0}$ to the self-energy. Thereafter, the Josephson current can be computed from Eq.~(\ref{eq:model.j}) and the approximate impurity Green function $\mc G=\tilde{\mc G}^{\la=0}$.

\section{Phase boundary}
\label{sec:pb}

\subsection{No direct coupling (\boldmath$t_d=0$)}
\label{sec:pb.td0}
For the simple quantum dot Josephson junction ($t_d=0$) it was previously demonstrated that the boundary between the singlet and doublet phases of the system is, even though roughly being governed by the ratio $\Delta/T_K$, an explicit function of all parameters of the system.\cite{gm,oguri1,oguri2,bcspaper} The latter are the Coulomb interaction $U$, the quantum dot energy $\epsilon$, the superconducting gap $\Delta$, the dot-lead hybridization $\Gamma$, and the phase difference $\phi$. According to Eq.~(\ref{eq:method.atom.pb}), the atomic-limit phase boundary for $t_d=0$ is determined by
\begin{equation}
U^2=4\epsilon^2+4\Gamma^2\cos^2(\phi/2),
\end{equation}
illustrating that an increase of either $U$ or $\phi$ drives the system towards the doublet phase, whereas a non-degenerate ground state is energetically favored the more $\epsilon$ is shifted away from particle-hole symmetry.\cite{gamma} Functional and numerical renormalization group calculations showed that the overall size of the doublet regime shrinks for $\Delta<\infty$, but all parameter dependencies of the phase boundary can still be understood in analogy to the case of $\Delta=\infty$.\cite{bcspaper} We will now demonstrate that the same holds true for the more complicated case of $t_d\neq0$.

\begin{figure*}[t]
\includegraphics[width=0.85\linewidth,clip]{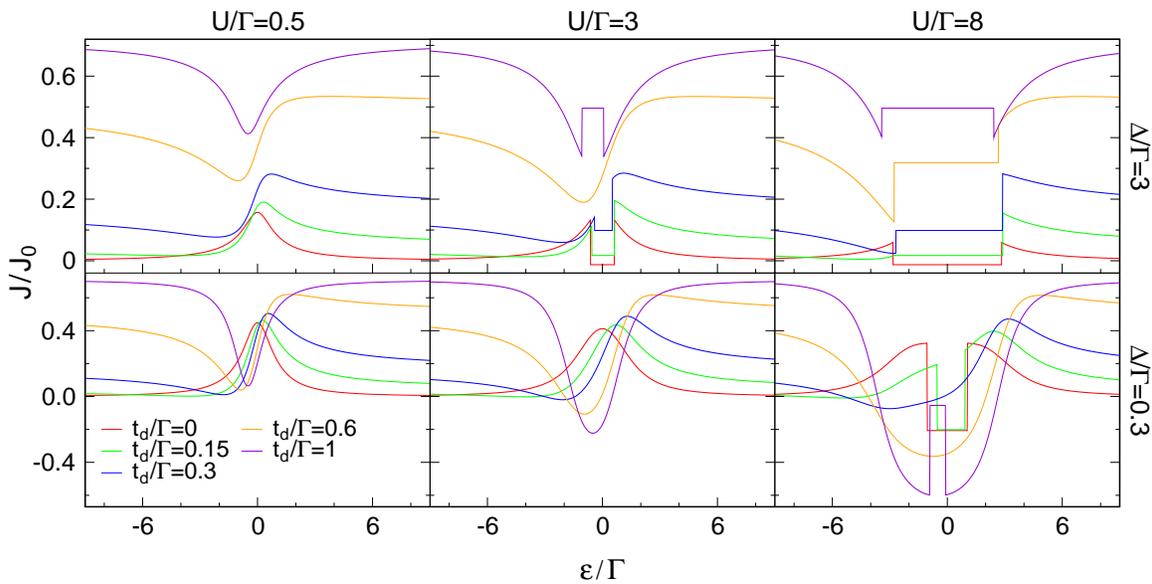}
\caption{(Color online) Josephson current $J$ (in units of $J_0=e\Delta/\hbar$) as a function of the impurity energy $\epsilon$ for constant $\phi=0.5\pi$ and $t_d/\Gamma=0$, $0.15$, $0.3$, $0.6$, and $1.0$ (from bottom to top at large $\epsilon$). The results were obtained from the FRG framework. In presence of a finite coupling $t_d$, $J(\epsilon)$ acquires a Fano-like lineshape analogous to the linear-response conductance of the ordinary Anderson model, and the non-monotonic phase boundary manifests as repeatedly appearing and disappearing discontinuities. In addition, one observes that the Josephson current can become negative in the singlet phase (see, e.g., the lower right panel which describes a singlet situation at intermediate and large $\epsilon$). The displayed behavior is generic for arbitrary phase differences $\phi$.}
\label{fig:j1}
\end{figure*}

\subsection{Aharonov-Bohm situation, \boldmath$\Delta=\infty$}
\label{sec:pb.atom}
In order to understand how a direct link between the superconductors affects the boundary of the singlet-doublet phase transition at $\Delta=\infty$, it is instructive to study the parameter dependence of the critical interaction strength $U_\tn{crit}$ for the case of particle-hole symmetry $\epsilon=0$ first. The quantity $U_\tn{crit}$ can be defined unambiguously, whereas, e.g., a critical coupling strength $t_d^\tn{crit}$ cannot due to the structure of Eq.~(\ref{eq:method.atom.pb}). For $\phi=0$ and $\phi=\pi$, one obtains
\begin{equation}
\frac{U^2_\tn{crit}}{4} =
\begin{cases}
\phantom{\tilde t_d^2\,}\Gamma^2/\,(1+\tilde t_d^2) & \phi=0 \\
\tilde t_d^2\,\Gamma^2 /\, (1-\tilde t_d^2)^2 & \phi=\pi
\end{cases}~,
\end{equation}
illustrating that there is a fundamental difference between both cases. At small $\phi$, the system is driven into the doublet phase if $t_d$ is increased [see, e.g., the curve for $\phi=0.2\pi$ in the inset to Fig.~\ref{fig:pb1}(b)]. In contrast, the phase boundary $U_\tn{crit}(t_d)$ depends non-monotonically on the direct coupling strength for $\phi=\pi$. At small values of $t_d$, $U_\tn{crit}$ increases quadratically, acquires a maximal value and finally falls off quadratically for large $t_d$ [see, e.g., the data for $\phi=0.8\pi$ in the inset to Fig.~\ref{fig:pb1}(b)]. Thus, a system which is initially in a doublet state can be driven into the singlet phase by increasing the coupling $t_d$ at fixed $U$, but eventually always re-enters the doublet phase.

Using Eq.~(\ref{eq:method.atom.pb}), one can show that the behavior of the phase boundary $U_\tn{crit}(t_d)$ for arbitrary $\phi$ is always qualitatively similar to the case of either $\phi=0$ or $\phi=\pi$. The onset of a non-monotonic dependence on $t_d$ occurs for $\phi\approx0.29\pi$, implying that one can expect to observe re-entrance behavior even if the phase difference cannot be controlled precisely. If the gate voltage is tuned away from the point of particle-hole degeneracy, the phase boundary exhibits an additional extremum [see Figs.~\ref{fig:pb1}(a,c)]. An additional minimum occurs for either large $\phi$ and arbitrary $\epsilon\neq0$ or small $\phi$ and $\epsilon<0$, whereas one observes an additional maximum for small $\phi$ and $\epsilon>0$. Since the critical value $U_\tn{crit}(t_d=0)$ is always larger than the asymptote $U_\tn{crit}(t_d\to\infty)$, the system can exhibit a total of three singlet-doublet phase transitions if the coupling strength $t_d$ is varied at fixed $U$ (and large $\phi$).

Due to the fact that the right-hand side of Eq.~(\ref{eq:method.atom.pb}) is a horizontally shifted quadratic function of the impurity energy $\epsilon$ [see FRG data for finite $\Delta<\infty$ in Fig.~\ref{fig:pb3}], the above-mentioned re-entrance behavior can also be observed by changing the impurity energy $\epsilon$ while fixing all other parameters. Tuning $\epsilon$ away from particle-hole symmetry can occasionally drive a system which is initially in a singlet state into the doublet and then ultimately back into the singlet phase. In contrast, the dependence of $U_\tn{crit}$ on the phase difference $\phi$ is always monotonous. At small $t_d$, a doublet ground state is favored if $\phi$ is increased, whereas the opposite holds for larger $t_d$ [see Fig.~\ref{fig:pb1}(a-c)]. For $\epsilon<0$ and $t_d\to\infty$, the system is again monotonously driven towards the doublet phase if $\phi$ is increased. One can analytically demonstrate that the crossover between the regimes of $\partial_\phi U_\tn{crit}\lessgtr0$ is characterized by values of $t_d$ where the phase boundary is completely independent of $\phi$.

\subsection{Aharonov-Bohm situation, arbitrary \boldmath$\Delta$}
\label{sec:pb.frg}
The truncated FRG scheme introduced in Sec.~\ref{sec:method.frg} allows for computing (an approximation to) the self-energy but does not yield the many-particle eigenstates of the system under consideration (in contrast, e.g., to the numerical renormalization group). Within this approach, the phase boundary is determined from discontinuities in the supercurrent with the understanding (based on the analytic treatment of the limit $\Delta=\infty$ as well as on NRG calculations at $t_d=0$\cite{oguri1}) that the groundstate is non-degenerate in the limit of small $U$.

For the simple quantum dot Josephson junction ($t_d=0$), comparison with NRG data illustrated that the approximate FRG scheme describes the phase boundary as well as the supercurrent both qualitatively and quantitatively at small to intermediate Coulomb correlations $U\lesssim8\Gamma$, whereas at larger $U$ qualitative features of both quantities are still captured correctly.\cite{bcspaper} Comparing FRG data for large $\Delta$ with the analytic result of Eq.~(\ref{eq:method.atom.pb}) shows that the former is also well-suited to tackle the problem at hand. The FRG reproduces all characteristics of the phase boundary at $t_d>0$ correctly [compare the insets of Figs.~\ref{fig:pb1}(a,b) with the main parts], only the size of the singlet phase is slightly overestimated. The latter tendency was already observed at $t_d=0$.

\begin{figure*}[t]
\includegraphics[width=0.85\linewidth,clip]{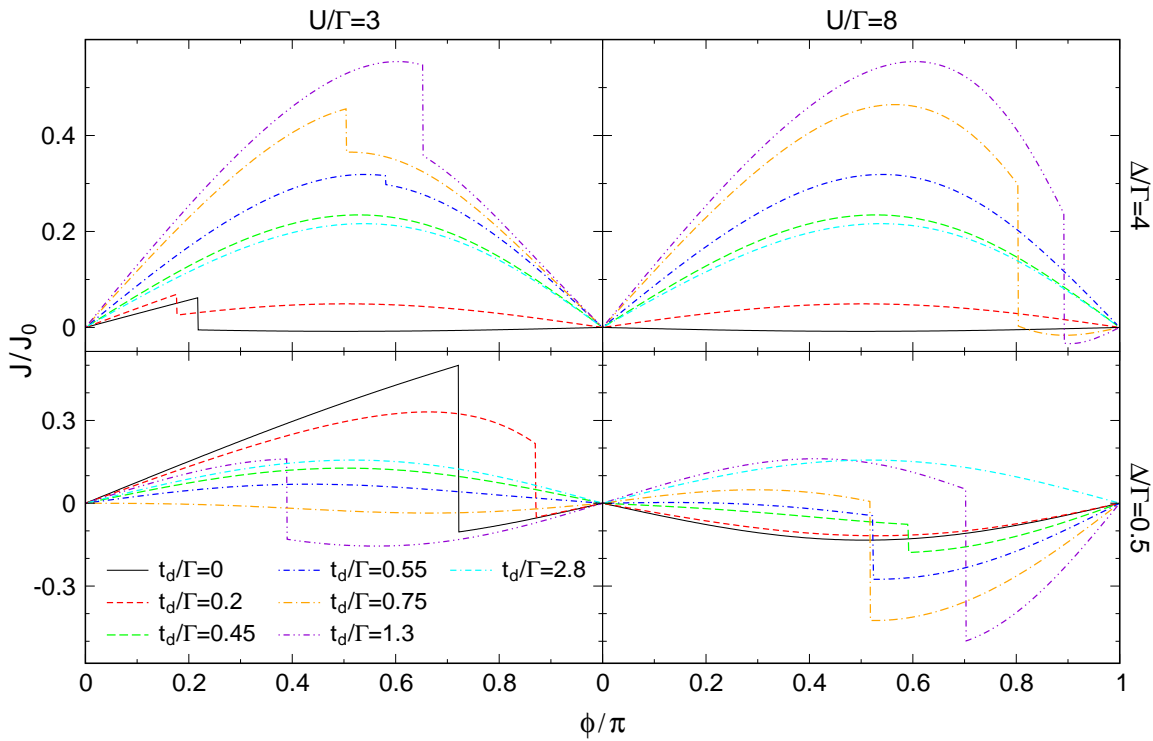}
\caption{(Color online) Josephson current $J$ (in units of $J_0=e\Delta/\hbar$) as a function of the phase difference $\phi$ for particle-hole symmetry $\epsilon=0$. Note that in the lower right panel the system is in a singlet state for $t_d\approx\Gamma$ and large $\phi$, illustrating that the current can become negative in this regime by virtue of the Coulomb interaction.}
\label{fig:j2}
\end{figure*}

FRG calculations at finite $\Delta<\infty$ demonstrate that all parameter dependencies of the phase boundary are similar to the case of $\Delta=\infty$, only the size of the doublet regime shrinks [see Figs.~\ref{fig:pb1}, \ref{fig:pb2} and \ref{fig:pb3}(a) for detailed comparisons of $U_\tn{crit}(t_d)$ and $U_\tn{crit}(\epsilon)$, respectively]. This is again consistent with results for the simple quantum dot Josephson junction ($t_d=0$).\cite{bcspaper} Since the FRG scheme, however, is approximative in $U$ but the critical interaction strength $U_\tn{crit}$ becomes large for small $\Delta$, it is reasonable to additionally study the phase boundary in terms of a different quantity. It turns out that a critical phase difference $\phi_\tn{crit}$ can always be defined unambiguously, and that the behavior of $\phi_\tn{crit}(t_d,U)$ for arbitrary $\Delta$ is similar to the atomic-limit solution [see Fig.~\ref{fig:pb3}(b)]. One can thus conclude that all parameter dependencies of the phase boundary can be understood from Eq.~(\ref{eq:method.atom.pb}), only the size of the doublet regime shrinks monotonously for finite $\Delta<\infty$.\cite{commenttk}

\section{Josephson current}
\label{sec:j}
In this Section, we present zero-temperature FRG results for the equilibrium supercurrent $J$ flowing through the Josephson junction in presence of a finite phase difference $\phi$ between the superconducting leads. According to Eq.~(\ref{eq:model.j}), this current can be interpreted to comprise of a `direct' and an `impurity' contribution. In contrast to the phase boundary, it is not determined solely by the dot Green function, rendering it impossible to derive an analytic result for $J$ in the limit $\Delta=\infty$. Thus, we focus exclusively on discussing FRG data for the Josephson current, again recalling that this framework was successfully benchmarked against numerical RG reference data for $t_d=0$.\cite{bcspaper}

In order to discuss how a direct link between the superconducting leads affects the supercurrent $J$, it is instructive to recall the simple quantum dot Josephson junction ($t_d=0$) first. For small $\Delta/T_K$, the system is in the singlet phase for all impurity energies $\epsilon$, and the current $J(\epsilon)$ exhibits a lineshape which resembles the linear-response conductance of the ordinary single impurity Anderson model (see the $t_d=0$ -- curves of Fig.~\ref{fig:j1}). In the opposite limit, $J(\epsilon)$ changes discontinuously at some critical value $\pm\epsilon_\tn{crit}$ as the system enters the doublet phase. The current becomes negative and almost independent of the impurity energy.\cite{stromfrg} In both cases, the evolution of $J(\epsilon)$ in presence of a finite link $t_d>0$ is Fano-like.\cite{zhang,oky} In addition, the non-monotonic dependence of the phase boundary on the coupling strength $t_d$ results in multiple singlet-doublet phase transitions manifesting as the appearance and disappearance of discontinuities of $J(\epsilon)$ (see Fig.~\ref{fig:j1}). One should particularly note that no matter how small $U$ or $\Delta$, the system will always enter the doublet phase in the limit of large hoppings $t_d$, provided that the impurity energy is not too large.

For the simple quantum dot Josephson junction ($t_d=0$), the supercurrent is always positive (negative) in the singlet (doublet) phase. Both does no longer necessarily hold in presence of a finite coupling $t_d$. Whereas it is rather intuitive that $J$ can become positive in the doublet regime due to the additional direct link (having in mind the ordinary Josephson junction where two superconductors are coupled by a hopping $t_d$ and $J>0$ holds for $0<\phi<\pi$), one can most notably also observe a negative current in the singlet phase,\cite{commentj} particularly at small BCS energy gaps $\Delta$ (see Fig.~\ref{fig:j1}). It is, however, imperative to point out that this is solely caused by the Coulomb interaction, and the supercurrent at $U=0$ (where the FRG becomes exact) always remains positive in the singlet phase. In contrast, Zhang (Ref.~\onlinecite{zhang}) obtains a negative singlet current yet in the non-interacting limit, rendering these results a priori highly questionable.

The Josephson current as a function of the phase difference $\phi$ displays the same characteristics as $J(\epsilon)$. Multiple phase transitions manifest as appearing and disappearing discontinuities of $J(\phi)$ (see Fig.~\ref{fig:j2}) and can be ultimately understood from the functional form of the atomic-limit phase boundary Eq.~(\ref{eq:method.atom.pb}). In addition, the current can become negative in the singlet phase in presence of both a direct coupling $t_d$ and finite Coulomb correlations. The actual form of $J(\phi)$ is rather complicated. It is displayed for various parameter sets in Fig.~\ref{fig:j2}.

\section{Conclusions and outlook}
\label{sec:outlook}
In this paper, we have investigated a quantum dot Josephson junction embedded within an Aharonov-Bohm environment. By analytically solving the atomic limit of large BCS gaps $\Delta=\infty$, we have shown that the low-energy physics of this system is governed by an interplay of two distinct (singlet and doublet) phases in complete analogy with the non-interferometric case where both superconductors are not coupled directly. The phase boundary, however, depends non-monotonically both on the coupling strength $t_d$ and the quantum dot energy $\epsilon$. By carrying out functional renormalization group calculations (which benchmark excellently against the atomic-limit result) at arbitrary $\Delta$, we have demonstrated that the overall size of the doublet regime shrinks monotonously with the gap size, but the functional form of the phase boundary always remains similar to the analytic expression derived at $\Delta=\infty$. Thus, even if all system parameters cannot be adjusted experimentally in a precisely controlled way, one can quite generally expect to observe re-entrance behavior within an interferometric quantum dot Josephson junction. At finite couplings $t_d$, the supercurrent $J(\epsilon)$ acquires a Fano-like lineshape analogous to the linear-response conductance of the ordinary Anderson model. Most importantly, we have shown that Coulomb correlations can cause $J$ to become negative in the singlet phase.

\section*{Acknowgdements}
We are grateful to A.~Oguri, T.~Novotn\'y, and J.~Paaske for fruitful discussions. This work was supported by the Deutsche Forschungsgemeinschaft via FOR 723.



\begin{thebibliography}{*}
\bibitem{zhangU0}
This was already pointed out in Ref.~\onlinecite{oky}.
\bibitem{commentkondo}
The FRG approximation introduced in Sec.~\ref{sec:method.frg} does not allow for computing the impurity spectral function. Thus, we cannot address the question whether there is actually a Kondo resonance or a BCS gap governing the low-energy behavior for the problem at hand. However, numerical renormalization group calculations for the simple quantum dot Josephson junction ($t_d=0$)\cite{oguri3} as well as for the non-superconducting setup ($\Delta=0$)\cite{fanokondo} showed that at sufficiently large $U$ the physics is crucially influenced by the Kondo effect. It is reasonable to assume that Kondo correlations are still present for the problem at hand.
\bibitem{zhangJ}
Focussing on the limit of small $\Delta/T_K$, this issue was previously raised in Ref.~\onlinecite{zhang}. However, the author fails to treat the non-interacting limit $U=0$ correctly,\cite{zhangU0} rendering his results a priori questionable.
\bibitem{gamma}
In the discussion of the phase boundary we implicitly assume that the hybridization strength $\Gamma$ is chosen as the unit of energy.
\bibitem{commenttk}
Since our FRG scheme turns out to be quantitatively reliable only up to some intermediate Coulomb interaction $U/\Gamma\approx8$ (where Kondo correlations start to become important), we cannot describe the singlet-doublet phase transition in the extreme limit of $\Delta/\Gamma\ll1$ (as this would require $T_K/\Gamma\ll1$), no matter which quantity is used to describe the phase boundary. However, for the simple quantum dot Josephson junction ($t_d=0$) it was demonstrated that generic physics shows up for small values of $T_K/\Delta$ (but not neccessarily extremely small $T_K$ itsself).\cite{oguri1,oguri2} Since we do not observe any qualitative changes apart from an overall increase of the size of the singlet phase in going from $\Delta=\infty$ to $\Delta/\Gamma\approx0.2$, it is reasonable to assume that the same holds for the problem at hand.
\bibitem{stromfrg}
For the simple quantum dot Josephson junction ($t_d=0$), it was previously observed that in the doublet phase the supercurrent obtained from the FRG is completely independent of both the two-particle interaction strength $U$ and the impurity energy $\epsilon$.\cite{bcspaper} Even though this is an artifact of the approximative approach, numerical RG calculations showed that physical properties only vary slightly in this regime.\cite{oguri2}
\bibitem{commentj}
Some generalized perturbative approach\cite{gm} could be employed in order to get some intuitive understanding why $J$ can become negative in the singlet phase. This is however, out of the scope of this paper.
\end{thebibliography}
\end{document}